\documentclass[11pt,twoside]{article}
\usepackage{asp2010}

\resetcounters

\begin{document}

\title{Prospects for cool white dwarf science from Pan-STARRS}
\author{Nigel Hambly,$^1$ Nick Rowell,$^2$ John Tonry,$^3$ Eugene Magnier$^3$ and Christopher Stubbs$^4$
\affil{$^1$Scottish Universities Physics Alliance (SUPA), Institute for Astronomy, School of Physics and Astronomy, University of Edinburgh, Royal Observatory, Blackford Hill, Edinburgh EH9~3HJ, UK}
\affil{$^2$Space Technology Centre, School of Computing, University of Dundee, Dundee DD1~4HN, UK}
\affil{$^3$Institute for Astronomy, University of Hawaii at Manoa, Honolulu, HI 96822, USA}
\affil{$^4$Department of Physics, Harvard University, Cambridge, MA 02138, USA}
}

\begin{abstract}
We discuss the prospects for new deep, wide--angle surveys of the Galactic cool white dwarf populations using data from Pan-STARRS: 
the Panoramic Survey Telescope \& Rapid Response System.
\end{abstract}

\section{Introduction}
Cool white dwarfs (WDs) provide the means to measure the star formation history
(see the article by Rowell elsewhere in these proceedings)
and age of the population to which they belong, as well as laboratories 
for the study of matter under extremes of pressure. Large samples 
cannot be culled from photometry alone --- cool WDs present colours
that are indistinguishable from those of the profusion of field F/G/K
dwarfs. However, using proper motion $\mu$ as a proxy for distance $d$
(where generally $\mu\propto 1/d$), we can employ the technique of
{\em reduced proper motion} (RPM) to distinguish the WDs from the much more 
luminous dwarfs provided we have a multi--epoch survey over a sufficiently
large time baseline to yield accurate proper motions.

To date, the widest angle studies have had to employ legacy Schmidt 
photographic surveys to provide multi--epoch imaging. The photographic 
plates are limited in their seeing (typically~2 to~3 arcsec) and depth 
(typically R~$\sim19$). Despite this, recent studies 
\citep[e.g.][]{2006AJ....131..571H} have 
produced large samples (up to $\sim10^4$) of cool WDs thereby enabling
determinations of the age of the disk population. However the
depth of the resulting luminosity functions is as yet insufficient to
determine the ages of the thick disk and spheroid of the Galaxy, and
the total space density of the coolest, oldest thick disk and spheroid
stars remains unknown \citep[e.g.][hereafter RH11]{2011MNRAS.417...93R}.

\section{Pan-STARRS}

The Panoramic Survey Telescope and Rapid Response 
System\footnote{\texttt http://pan-starrs.ifa.hawaii.edu/public/} 
prototype `PS1' --- see \cite{2012AJ....143..142M} and references therein --- is now well into
it's 3~year wide--angle observing campaign (the so--called `$3\pi$' survey), 
aiming to survey 75\% of the sky 
in the 5 grizy$_{\rm P1}$ passbands at least 4 times per year yielding a
multi--epoch survey of stunning proportions. With a typical seeing of
$\sim1$ arcsec, a depth surpassing the original SDSS, and 
near--infrared coverage that was impossible using photographic plates,
Pan-STARRS will provide a rich hunting ground for cool WDs. At the time
of writing, the $3\pi$ survey is essentially completely covered in all
passbands with the expected number of individual epochs for this stage in the
project; overall observational completeness stands at 59\%. A fully world--public
release of PS1 data is presently scheduled for the end of 2014; in the meantime,
data access is restricted to scientists within those institutes that are
part of the PS1 Science Consortium.

\section{Depth and completeness}

Figure~\ref{nmhists} shows number--magnitude histograms for a 200~square degree equatorial strip
($10^\circ<\alpha<35^\circ$ and $-4.0^\circ<\delta<+4.0^\circ$)
across the South Galactic Cap from the extant Pan-STARRS $3\pi$ survey.
To be counted here, detections must be present in at least gri$_{\rm P1}$ with at least 10
individual epochs. We compare also with the Rowell \& Hambly survey sample by
identifying the same objects in those catalogues employing the closest
corresponding passbands (BRI). The increase in depth afforded
by Pan-STARRS is clear, as is the general incompleteness in the photographic
data (up to $\sim50$\%) owing to the poorer seeing.

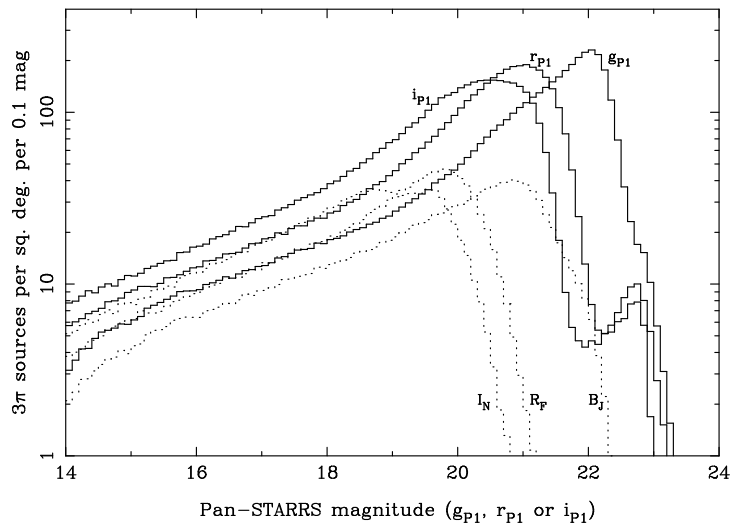
\articlefigure[scale=0.4,angle=270]{hambly_poster_fig1.ps}{nmhists}{Number-magnitude
histograms of sources in the extant $3\pi$ survey (solid lines labelled by passband) compared with similar from the
SuperCOSMOS legacy Schmidt photographic survey (dotted lines labelled by passband).}

\section{Reduced Proper Motion diagrams}

Proper motions from the full 3yr Pan-STARRS $3\pi$ survey are of course
not yet available, but we can get a good idea of the potential gains
by combining the new photometry with the legacy photographic plate proper
motions. Figure~\ref{rpm} shows 3 RPM diagrams: on the left is that from
the photographic plate survey data alone for the SGC subsample of the RH11
survey (R~$<19.75$); the middle panel shows the same 
objects but using the Pan-STARRS photometry; finally on the right we show the
RPM diagram combining Pan-STARRS photometry with SuperCOSMOS proper motions
without the R=19.75 cutoff. The tighter subdwarf sequence and cleaner separation
of the WDs is clear when using Pan-STARRS, as is the enhanced depth (despite
the necessity of employing still the photographic plate astrometry).

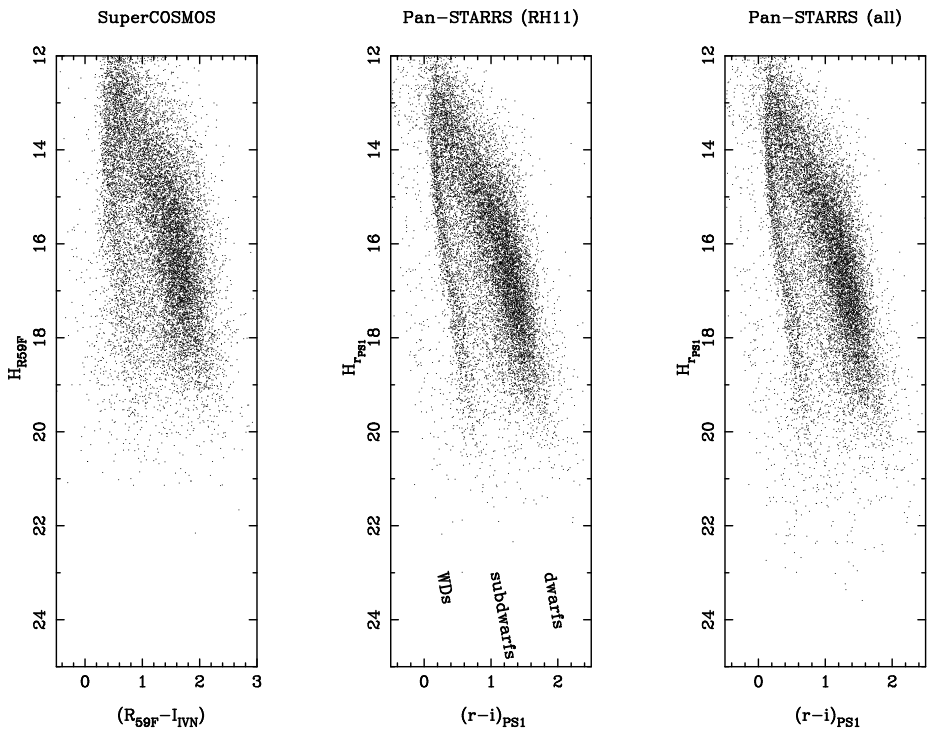
\articlefigure[scale=0.5,angle=270]{hambly_poster_fig2.ps}{rpm}{RPM diagrams for
the \citeauthor{2011MNRAS.417...93R} sample using the photographic photometry (left panel); the same
sample but using PS1 photometry (middle panel); and a sample employing the photographic
astrometry and PS1 photometry without the R~$<19.75$ cut employed by RH11 (right panel).}

\section{Discussion}

\cite{2012ApJ...745...42T} present new results from the Pan-STARRS Medium Deep Survey;
however we suggest that it is the $3\pi$ survey that will sample the greatest volume for the coolest WDs,
and will discover many objects amenable to spectroscopic follow-up on 4~to~8m--class
facilities. The prospective numbers of cool WDs that will be detected in the $3\pi$
survey can be calculated using simple scaling arguments from the RH11 sample.
Volume sampled for a uniformly distributed population (e.g.~the local
spheroid population) will go as $d^3$ for distance $d$ limited by magnitude
($d_{\rm P1}/d_{\rm RH11}=10^{({\rm r}_{\rm P1}-{\rm r}_{\rm RH11})/5}$) 
where ${\rm r}_{\rm P1}$ is the magnitude limit for a sample drawn from the PS1 $3\pi$ survey with
proper motion characteristics similar to RH11. 
They found that RPM discriminates usefully with proper motions as low as $5\sigma_\mu$ and employed
a magnitude--dependent lower proper motion limit typically between $60<5\sigma_\mu<100$~mas/yr 
at the limit of the SuperCOSMOS survey data (their Figure~2). To estimate PS1 proper motion precision, 
we note the \cite{1985MNRAS.214..575I} rule--of--thumb which states that in units of the scale size of well
sampled faint point sources in uniform background noise (0.6~arcsec for PS1 corresponding to 
half--width at half--maximum), centroiding precision $\sigma_x$ is equal to relative flux precision $\sigma_f/f$. 
Hence $5\sigma_f$ detections will have centroids accurate to $\sigma_x\approx0.12$~arcsec, 
and given $N=60$ measurements over $\Delta t=3.5$~yr, total proper motion precision will be typically 
$\sigma_\mu=\surd2\times(\sigma_x/\Delta t)(12/N)^{0.5}\approx22$~mas~yr$^{-1}$ where the factor~12 comes
from the variance of a uniform distribution. For a lower proper motion limit comparable to RH11, 
say $5\sigma_\mu=80$~mas/yr, we require $\sigma_\mu=16$~mas~yr$^{-1}$ which corresponds to $\sigma_x\sim87$~mas. 
This should be achieved a factor 1.4 (or 0.35~mag) brighter than the $5\sigma_f$ detection limit, 
the latter corresponding to 50\% completeness at $r_{\rm P1}\sim21.6$ according to Figure~\ref{nmhists};
hence we choose $r_{\rm P1}\sim21.25$. Noting the 50\% incompleteness in RH11 and that r$_{\rm RH11}=19.75$, 
the number of PS1 spheroid WDs will be $\sim 2\times2^3\approx16$ times
higher. For the thin disk component, scale height effects reduce the
distance exponent to~2, so the factor increase is $2\times2^2\approx8$; the thick
disk increase will be somewhere  between the two. Not only will the sample be significantly 
larger, but the availability of 5--colour photometry will greatly improve the RH11 analysis
which was limited to 50\% errors in photometric distances from 3--colour photographic
photometry.


\acknowledgements The Pan-STARRS1 Surveys (PS1) have been made possible through contributions of the Institute for Astronomy, 
the University of Hawaii, the Pan-STARRS Project Office, the Max-Planck Society and its participating institutes, 
the Max Planck Institute for Astronomy, Heidelberg and the Max Planck Institute for Extraterrestrial Physics, Garching, 
The Johns Hopkins University, Durham University, the University of Edinburgh, Queen's University Belfast, 
the Harvard-Smithsonian Center for Astrophysics, the Las Cumbres Observatory Global Telescope Network Incorporated, 
the National Central University of Taiwan, the Space Telescope Science Institute, 
and the National Aeronautics and Space Administration under Grant No. NNX08AR22G issued through the 
Planetary Science Division of the NASA Science Mission Directorate.

\bibliography{hambly_poster}

\end{document}